\documentclass[sigconf]{acmart}

\usepackage{xcolor, graphicx}
\usepackage{color, colortbl}
\usepackage{tcolorbox}
\usepackage[export]{adjustbox}
\usepackage{subcaption}
\usepackage{multirow}
\usepackage{ulem}
\usepackage{enumitem}

\newcommand\BibTeX{B\textsc{ib}\TeX}
\AtBeginDocument{%
  \providecommand\BibTeX{{%
    \normalfont B\kern-0.5em{\scshape i\kern-0.25em b}\kern-0.8em\TeX}}}

\copyrightyear{2021}
\acmYear{2021}
\setcopyright{acmcopyright}\acmConference[SIGIR '21]{Proceedings of the 44th International ACM SIGIR Conference on Research and Development in Information Retrieval}{July 11--15, 2021}{Virtual Event, Canada}
\acmBooktitle{Proceedings of the 44th International ACM SIGIR Conference on Research and Development in Information Retrieval (SIGIR '21), July 11--15, 2021, Virtual Event, Canada}
\acmPrice{15.00}
\acmDOI{10.1145/3404835.3463080}
\acmISBN{978-1-4503-8037-9/21/07}

\acmSubmissionID{sp1604}

\begin{document}
\fancyhead{}

\title{Understanding the Role of Affect Dimensions in Detecting Emotions from Tweets: A Multi-task Approach}

\author{Rajdeep Mukherjee}
\affiliation{\institution{IIT Kharagpur, India}}
\email{rajdeep1989@iitkgp.ac.in}

\author{Atharva Naik}
\authornote{Both authors contributed equally to this research.}
\affiliation{\institution{IIT Kharagpur, India}}
\email{atharvanaik2018@iitkgp.ac.in}

\author{Sriyash Poddar}
\authornotemark[1]
\affiliation{\institution{IIT Kharagpur, India}}
\email{poddarsriyash@iitkgp.ac.in}

\author{Soham Dasgupta}
\affiliation{\institution{MAIS Bangalore, India}}
\email{sohamdasgupta91@gmail.com}

\author{Niloy Ganguly}
\affiliation{\institution{IIT Kharagpur, India}}
\email{niloy@cse.iitkgp.ac.in}

\renewcommand{\shortauthors}{Rajdeep Mukherjee, et al.}

\begin{abstract}
  We propose \textbf{VADEC}, a multi-task framework that exploits the correlation between the \textit{categorical} and \textit{dimensional} models of emotion representation for better subjectivity analysis. Focusing primarily on the effective detection of emotions from tweets, we jointly train \textit{multi-label emotion classification} and \textit{multi-dimensional emotion regression}, thereby utilizing the inter-relatedness between the tasks. Co-training especially helps in improving the performance of the \textit{classification} task as we outperform the strongest baselines with 3.4\%, 11\%, and 3.9\% gains in \textit{Jaccard Accuracy}, \textit{Macro-F1}, and \textit{Micro-F1} scores respectively on the \textit{AIT} dataset \cite{Mohammad2018SemEval2018T1}. We also achieve state-of-the-art results with 11.3\% gains averaged over six different metrics on the \textit{SenWave} dataset \cite{Yang2020SenWaveMT}. For the \textit{regression} task, \textit{VADEC}, when trained with \textit{SenWave}, achieves 7.6\% and 16.5\% gains in \textit{Pearson Correlation} scores over the current state-of-the-art on the \textit{EMOBANK} dataset \cite{buechel-hahn-2017-emobank} for the Valence (V) and Dominance (D) affect dimensions respectively. We conclude our work with a case study on COVID-19 tweets posted by Indians that further helps in establishing the efficacy of our proposed solution.
\end{abstract}




\maketitle

\section{Introduction}\label{intro}
%Emotions are an integral part of any communication as they reflect how we perceive the world around us. 
With the proliferation of social media, as more and more people express their opinions online, detecting human emotions from their written narratives, especially tweets has become a crucial task given its widespread applications in e-commerce, public health monitoring, disaster management, etc. \cite{mohammad-kiritchenko-2018-understanding, Mohammad2018SemEval2018T1}. \if 0 Owing to the complexity of human psychology, researchers over the years have proposed several models to represent emotions. While\fi \textit{Categorical} models of emotion representation such as Plutchik's \textit{Wheel of Emotion} \cite{PLUTCHIK19803} or Ekman's \textit{Basic Emotions} \cite{Ekman1992AnAF} classify affective states into discrete categories (joy, anger, etc.). \textit{Dimensional} models on the other hand describe emotions relative to their fundamental dimensions. Russel and Mehrabian's \textit{VAD} model \cite{RUSSELL1977273} for instance interprets emotions as points in a 3-D space with \textit{Valence} (degree of pleasure or displeasure), \textit{Arousal} (degree of calmness or excitement), and \textit{Dominance} (degree of authority or submission) being the three orthogonal dimensions. Accordingly, the literature on text-based emotion analysis can be broadly divided into \textit{coarse-grained classification} systems \cite{yu-etal-2018-improving, huang2019seq2emo, Jabreel2019ADL, Fei_Zhang_Ren_Ji_2020, pg-cods} and \textit{fine-grained regression} systems \cite{yu-etal-2015-predicting, wang-etal-2016-dimensional, preotiuc-pietro-etal-2016-modelling, zhu-etal-2019-adversarial}. Although a \textit{coarse-grained} approach is better-suited for the task of detecting emotions from tweets as observed in \cite{beuchel_hahn_ecai2016}, prior works fail to exploit the direct correlation between the two models of emotion representation for finer interpretation. %subjectivity analysis. 
We \if 0 on the other hand\fi utilize the better representational power of \textit{dimensional} models \cite{beuchel_hahn_ecai2016} to improve the \textit{emotion classification} performance by proposing \textbf{VADEC} that jointly trains \textit{multi-label emotion classification} and \textit{multi-dimensional emotion regression} in a multi-task framework.

Multi-task learning \cite{Caruana2004MultitaskL} has been successfully used across a wide spectrum of NLP tasks including emotion analysis \cite{all_in_one, zhu-etal-2019-adversarial}. \if 0 in order to learn better generalization capabilities by training multiple related tasks together.\fi While \textit{AAN} \cite{zhu-etal-2019-adversarial} takes an adversarial approach to learn discriminative features between two emotion dimensions at a time, \if 0 and hence requires three different models to be trained for predicting the VAD scores.\fi \textit{All\_In\_One} \cite{all_in_one} proposes a multi-task ensemble framework to learn different configurations of tasks related to coarse- and fine-grained sentiment and emotion analysis. However, none of the methods combine the supervisions from VAD and categorical labels. Our proposed framework (Section \ref{sec:model}) consists of a \textbf{classifier} module that is trained for the task of multi-label emotion classification, %using supervisions from the \textit{AIT} dataset \cite{Mohammad2018SemEval2018T1}, 
and a \textbf{regressor} module that co-trains the regression tasks corresponding to the V, A, and D dimensions. %using supervisions from the \textit{EMOBANK} dataset \cite{buechel-hahn-2017-emobank}. 
Owing to the unavailability of a common annotated corpus, the two tasks are trained using supervisions from their respective benchmark datasets (reported in Section \ref{subsec:dataset}), which further justifies the utility of our proposed multi-task approach.

\textit{VADEC} learns better shared representations by jointly training the two modules, that especially help in improving the performance of the \textit{classification} task, thereby achieving state-of-the-art results on the \textit{AIT} \cite{Mohammad2018SemEval2018T1} and \textit{SenWave} \cite{Yang2020SenWaveMT} datasets (Section \ref{subsec:eval_classification}). For the \textit{regression} task, we achieve SOTA results on the \textit{EMOBANK} dataset \cite{buechel-hahn-2017-emobank} for \textit{V} and \textit{D} dimensions (Section \ref{subsec:eval_regression}). We conclude our work with a detailed case study in Section \ref{subsec:case_study}, where we apply our trained multi-task model to detect and analyze the changing dynamics of Indian emotions towards the COVID-19 pandemic from their tweets. We discover the major factors contributing towards the various emotions and find their trends to correlate with real-life events.
\section{\textit{VADEC} Architecture}\label{sec:model}
\begin{figure}
    \centering
    \begin{tabular}{c}
        \includegraphics[width=0.65\columnwidth]{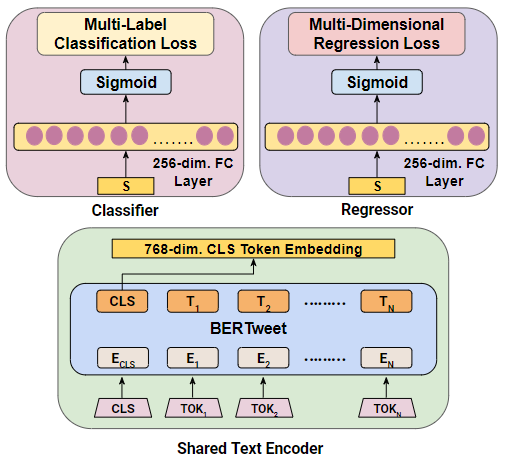}
        \\
        \includegraphics[width=0.65\columnwidth]{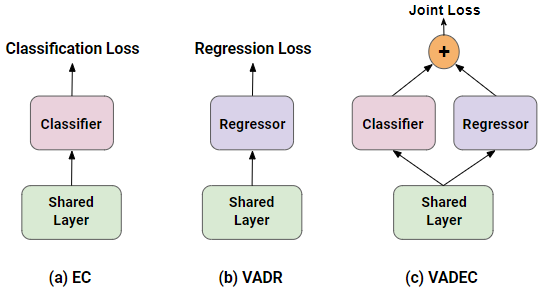}
    \end{tabular}
    \caption{Components and Model Architecture: Pre-trained BERTweet serves as the \textit{Shared Text Encoder} between the \textit{Classifier} and \textit{Regressor} modules.\if 0 768-dim. $[CLS]$ token embedding of the text is sent as input to the \textit{Classifier} and \textit{Regressor} modules.\fi (a) \textit{EC} and (b) \textit{VADR} respectively represent the Multi-label Emotion \textit{Classifier} and Multi-dimensional Emotion \textit{Regressor} when trained individually. (c) \textit{VADEC} represents our Multi-Task Affect Classifier that co-trains the two modules by optimizing the joint loss.}
    \label{fig:model}
    % \vspace{-1em}
\end{figure}

Figure \ref{fig:model} illustrates the architecture of \textit{VADEC}, that jointly trains a multi-label emotion \textit{classifier} and a multi-dimensional emotion \textit{regressor} with supervision from their respective datasets. Since we primarily focus on detecting emotions from tweets, we use \textit{BERTweet} \cite{bertweet} \if 0, the first public large scale language model pre-trained with 850M English tweets and 5M COVID-19 related tweets,\fi to serve as our text-encoder. It is shared by the two modules and is hereby referred to as the \textit{shared layer}. The 768-dim. $[CLS]$ token embedding of the sentence/tweet obtained from BERTweet is first passed through a fully connected (FC) layer with 256 neurons in both the modules respectively. The \textit{classifier} passes this intermediate representation through another FC layer with 11 output neurons, each activated using \textit{Sigmoid} with a threshold of 0.5 to predict the presence/absence of one of the 11 emotion categories. \textit{Binary Cross-Entropy} (BCE) with L2-norm regularization is used as the loss function, hereby referred to as the \textit{EC\textsubscript{Loss}}. Similarly, the \textit{regressor} passes the 256-dim. intermediate representation through an FC layer with 3 output neurons (with \textit{Sigmoid} activation) corresponding to the V, A and D dimensions. It then jointly optimizes the \textit{Mean Squared Error} (MSE) loss of all three dimensions, hereby referred to as the \textit{VADR\textsubscript{Loss}}. \textit{VADEC} jointly trains the two modules by optimizing the following multi-task objective:
\begin{equation}\label{eq:mtl_obj}
    VADEC\textsubscript{Loss} = \lambda \cdot \textit{EC\textsubscript{Loss}} + (1 - \lambda) \cdot \textit{VADR\textsubscript{Loss}}
\end{equation}
Here, $ \lambda $ represents a balancing parameter between the two losses. The weighted joint loss backpropagates through the \textit{shared layer}, thereby fine-tuning the \textit{BERTweet} parameters end-to-end.

\section{Results and Discussion}\label{sec:results}
\subsection{Datasets}\label{subsec:dataset}
For our experiments, we consider \textit{EMOBANK}, a VAD dataset, and two categorical datasets, \textit{AIT} and \textit{SenWave} as described below:
\begin{itemize}[leftmargin=*]
    \item \textbf{EMOBANK} (Buechel and Hahn \cite{buechel-hahn-2017-emobank}) : A collection of around 10k English sentences from multiple genres (8,062 for training, and 1K sentences each for validation and testing), each annotated with continuous scores (in the range of 1 to 5) for \textit{Valence}, \textit{Arousal}, and \textit{Dominance} dimensions of the text.
    \item \textbf{AIT} (Mohammad et al. \cite{Mohammad2018SemEval2018T1}) : Created as part of SemEval 2018 Task 1: ``Affect in Tweets'', it consists of 10,983 English tweets (6,838 for training, 886 for validation, 3,259 for testing), each with labels denoting the presence/absence of a total of 11 emotions. \if 0, viz. \textit{anger, anticipation, disgust, fear, joy, love, optimism, pessimism, sadness, surprise}, and \textit{trust}.\fi
    \item \textbf{SenWave} (Yang et al. \cite{Yang2020SenWaveMT}) : Till date the largest fine-grained annotated COVID-19 tweets dataset consisting of 10K English tweets (8K for training, and 1K each for validation and testing), each with corresponding labels denoting the presence/absence of 11 different emotions specific to COVID-19.
\end{itemize}
	
\subsection{Experimental Setup}\label{subsec:setup}
For all our model variants, we perform extensive experiments with different sets of hyper-parameters and select the best set w.r.t. lowest validation loss. %(\textit{BCE} loss for \textit{EC}, \textit{MSE} loss for \textit{VADR}, and \textit{joint loss} for \textit{VADEC} by empirically setting $\lambda$ in Eq. \ref{eq:mtl_obj} to 0.5). 
Before evaluating the performance on the test set, we combine the training and validation data and re-train the models with the best obtained set of hyper-parameters (learning rate $= 2e-5$, weight decay $= 0.01$, $\lambda$ = 0.5, and no. of epochs $= 5$ for \textit{VADEC}). For the \textit{regression} task, the outputs of \textit{Sigmoid} activation at each of the three output neurons are suitably scaled before calculating the \textit{MSE} loss since the ground-truth VAD scores are in the range of 1-5. As \textbf{model ablations}, we investigate the role played by features derived from affect lexicons by additionally appending a 194-dim. \textit{Empath}\footnote{\url{https://github.com/Ejhfast/empath-client}} \cite{empath} feature vector to the intermediate representations learnt by our model variants to be used for final predictions. Parameters of our \textit{shared encoder} are initialized with pre-trained model weights (\textit{roberta-base} for RoBERTa, and \textit{bertweet-base} for BERTweet) from the \textit{HuggingFace Transformers} library \cite{Wolf2019HuggingFacesTS}. Other model parameters are randomly initialized. All our model variants are trained end-to-end with \textit{AdamW} optimizer \cite{Loshchilov2019DecoupledWD} on Tesla P100-PCIE (16GB) GPU. We additionally ensure the reproducibility of our results and make our code repository \footnote{\url{https://github.com/atharva-naik/VADEC}} publicly accessible.

\subsection{Evaluating Emotion Classification}\label{subsec:eval_classification}
We first discuss the comparative results of our model variants and ablations on the \textbf{AIT} dataset. We then respectively report our state-of-the-art results achieved on the \textit{AIT} and the \textbf{SenWave} datasets.

\subsection*{AIT Dataset}
As \textbf{metrics} we use \textit{Jaccard Accuracy}, \textit{Macro-F1}, and \textit{Micro-F1} \cite{Mohammad2018SemEval2018T1}. 
Among recent \textbf{baselines}: (i) \textbf{BERTL} (Park et al. \cite{VAD-SOTA}) denotes the scores obtained by fine-tuning BERT-Large \cite{Devlin2019BERTPO} on the \textit{AIT} dataset, and (ii) \textbf{NTUA-SLP} (Baziotis et al. \cite{Baziotis2018NTUASLPAS}) represents the winning entry for this (sub)task of SemEval 2018 Task 1 \cite{Mohammad2018SemEval2018T1}, where the authors take a transfer learning approach by first pre-training their Bi-LSTM  architecture, equipped with multi-layer self attentions, on a large collection of general tweets and the dataset of SemEval 2017 Task 4A, before fine-tuning their model on this dataset. Among our \textbf{model variants and ablations}: (i) \textbf{EC} represents our \textit{classifier} module, when trained as a single task (Fig. \ref{fig:model}a), (ii) \textbf{EC\textsubscript{RoBERTa}} uses \textit{RoBERTa} \cite{Liu2019RoBERTaAR} instead of \textit{BERTweet} as the shared layer.
 
From Table \ref{tab:ec_comparison_ait}, \textit{NTUA-SLP} surprisingly outperforms \textit{BERTL} (on \textit{Jac. Acc.} and \textit{Micro-F1}), a heavier model with 336M parameters. \textit{EC} (trained with \textit{BERTweet}) comfortably beats \textit{EC\textsubscript{RoBERTa}} demonstrating the better efficacy of \textit{BERTweet} in learning features from tweets. The sparse \textit{Empath} feature vectors do not however add any value to the rich 768-dim. contextual representations learnt using BERT-based methods. We obtain our \textbf{best results with \textit{VADEC}}, with respectively \textbf{3.4\%, and 3.9\% gains in \textit{Jacc. Acc.}, and \textit{Micro-F1}} over \textit{NTUA-SLP}, and \textbf{11\% gain} in \textbf{\textit{Macro-F1}} over \textit{BERTL}.
\if 0 Our model is thus better equipped to handle imbalanced class distribution, a common issue with most real-life classification problems.\fi

\subsection*{SenWave Dataset}
%The  ablation study performed on AIT already proves the superiority of VADEC over its different variants. So here we undertake a detail comparison with the technique proposed by the authors of \textbf{\textit{SenWave}} dataset.
Considering the superior performance of \textit{VADEC} over all its model variants and ablations from Table \ref{tab:ec_comparison_ait}, here we directly compare the results of \textit{VADEC}, re-trained with \textit{SenWave} \cite{Yang2020SenWaveMT}, with the ones reported by the authors of \cite{Yang2020SenWaveMT}, serving as the only available \textbf{baseline} on this dataset. Following \cite{Yang2020SenWaveMT}, we use \textit{Label Ranking Average Precision} (LRAP), \textit{Hamming Loss}, and \textit{Weak Accuracy} (Accuracy) as \textbf{metrics} in addition to the ones reported in Table \ref{tab:ec_comparison_ait}. %besides \textit{Jaccard Accuracy}, \textit{Macro-F1}, and \textit{Micro-F1}.
As observed from Table \ref{tab:ec_comparison_sw}, \textbf{\textit{VADEC} achieves SOTA} by outperforming the baseline scores with \textbf{11.3\%} performance gain averaged over \textbf{all 6 metrics}.

Overall, our results from Tables \ref{tab:ec_comparison_ait} and \ref{tab:ec_comparison_sw} demonstrate the advantage of  utilizing the VAD supervisions for improving the performance of the multi-label emotion classification task.

\subsection{Evaluating Emotion Regression}\label{subsec:eval_regression}
Pearson Correlation Coefficient \textit{r} is used as the evaluation \textbf{metric} for this task. All the models are evaluated on the \textit{EMOBANK} dataset. Among recent \textbf{baselines}: (i) \textbf{AAN} (Zhu et al. \cite{zhu-etal-2019-adversarial}) \if 0 represents an Adversarial Attention Network that\fi employs adversarial learning between two attention layers to learn discriminative word weight parameters for scoring two emotion dimensions at a time. The authors report the VAD scores for all 6 domains and 2 perspectives of \textit{EMOBANK}. For comparison, we use their highest correlation score for each dimension, (ii) \textbf{All\_In\_One} (Akhtar et al. \cite{all_in_one}) represents a multi-task ensemble framework which the authors use for learning four different configurations of multiple tasks related to emotion and sentiment analysis, (iii). \textbf{SVR-SLSTM} (Wu et al. \cite{WU201930}) represents a semi-supervised approach using variational autoencoders to predict the VAD scores, and (iv). \textbf{BERTL (EB $\leftarrow$ AIT)} \cite{VAD-SOTA}, the current state-of-the-art, fine-tunes BERT-Large \cite{Devlin2019BERTPO} on the \textit{AIT} dataset to predict VAD scores by means of minimizing EMD distances between the predicted VAD distributions and sorted categorical emotion distributions as a proxy for target VAD distributions. For comparison, we use their reported scores obtained upon further fine-tuning their best-trained model on the \textit{EMOBANK} corpus. Our \textbf{model variants} include (i) \textbf{VADR} which represents our \textit{regressor} module, when trained as a single task (Fig. \ref{fig:model}b), (ii) \textbf{VAD\textsubscript{RoBERTa}}, an ablation where we experiment with \textit{RoBERTa} as the shared layer, (iii) \textbf{VADEC (AIT)}, and (iv) \textbf{VADEC (SenWave)} representing the scores of our multi-task model when trained respectively with the \textit{AIT} and \textit{SenWave} datasets.

From Table \ref{tab:vad_comparison}, \textit{VADR\textsubscript{RoBERTa}} shows the highest correlation ($0.511$) on the \textit{D} dimension. \textit{VADR} (w/ BERTweet) however outperforms \textit{VADR\textsubscript{RoBERTa}} on the other two dimensions. %whose performance further degrades when we additionally use \textit{Empath} derived features to train the model. 
Contrary to our observations in the \textit{classification} task, co-training does not help in improving the performance of the \textit{regression} task, as can be confirmed from the results of \textit{VADEC} (AIT) and \textit{VADR}. \if 0 Similar observations are also reported in \cite{VAD-SOTA}.\fi Although we are outclassed by \textit{BERTL (EB $\leftarrow$ AIT)} on the \textit{A} dimension, \textit{VADEC} (AIT) comfortably outperforms \textit{BERTL (EB $\leftarrow$ AIT)} on the \textit{V} and \textit{D} dimensions. \textbf{VADEC (SenWave)} further outclasses both \textit{VADEC} (AIT) and \textit{BERTL (EB $\leftarrow$ AIT)} on \textit{V} and \textit{D} with \textbf{7.6\% and 16.5\% gains} respectively. To conclude, although joint-learning does not help the \textit{regression} task as much as it helps in improving the \textit{classification} performance (which in fact is our main objective), we still achieve noticeable improvements in majority of emotion dimensions. 

\begin{table}
	\caption{Comparative Results on the \textit{AIT}. Results of \textit{VADEC} are statistically significant than \textit{EC} with 95\% conf. interval.}
	\label{tab:ec_comparison_ait}
	\vspace{-1em}
	\centering
	\begin{adjustbox}{max width=\columnwidth}
	\begin{tabular}{|lccc|}
		\hline
		\textbf{Methods} & \textbf{Jaccard Acc.} & \textbf{F1-Macro} & \textbf{F1-Micro} \\ \hline
		
		BERTL \cite{VAD-SOTA} & 0.572 & 0.534 & 0.697 \\
		NTUA-SLP \cite{Baziotis2018NTUASLPAS} & 0.588 & 0.528 & 0.701 \\
		\hline
		\hline
		EC\textsubscript{RoBERTa} & 0.592 & 0.570 & 0.712 \\
		\quad w/ Empath & 0.585 & 0.562 & 0.706 \\
		\hline
		EC & 0.605 & 0.581 & 0.723 \\
		\quad w/ Empath & 0.602 & 0.570 & 0.720 \\
		\hline
		VADEC & \textbf{0.608} & \textbf{0.593} & \textbf{0.728}\\
		\hline
		\hline
		\textit{Significance T-Test (p-values)} & 0.029 & - & - \\
		\hline
	\end{tabular}
	\end{adjustbox}
\end{table}

\begin{table}
    \vspace{-0.5em}
    \centering
	\caption{Comparative Results on the \textit{SenWave} dataset.}
	\label{tab:ec_comparison_sw}
	\vspace{-1em}
	\begin{adjustbox}{max width=\columnwidth}
	\begin{tabular}{|l|c|c|c|c|c|c|}
		\hline
		\textbf{Methods} & \textbf{Accuracy} & \textbf{Jac. Acc.} & \textbf{F1-Macro} & \textbf{F1-Micro} & \textbf{LRAP} & \textbf{Ham. Loss} \\ 
		\hline
		SenWave \cite{Yang2020SenWaveMT} & 0.847 & 0.495 & 0.517 & 0.573 & 0.745 & 0.153 \\
		\hline
		VADEC & \textbf{0.877} & \textbf{0.560} & \textbf{0.563} & \textbf{ 0.620} & \textbf{0.818} & \textbf{0.123} \\
		\hline
	\end{tabular}
	\end{adjustbox}
\end{table}

\begin{table}
	\vspace{-0.5em}
	\caption{Comparison of Pearson Correlation (r-values) for the emotion regression task on the \textit{EMOBANK} (EB) dataset.}
	\label{tab:vad_comparison}
	\vspace{-1em}
	\centering
	\begin{adjustbox}{max width=\columnwidth}
	\begin{tabular}{|lccc|}
		\hline
		\textbf{Methods} & \textbf{Valence (V)} & \textbf{Arousal (A)} & \textbf{Dominance (D)} \\ 
		\hline
        AAN \cite{zhu-etal-2019-adversarial} & 0.424 & 0.351 & 0.265 \\
		All\_In\_One \cite{all_in_one} & 0.635 & 0.375 & 0.277 \\
		SRV-SLSTM \cite{WU201930} & 0.620 & 0.508 & 0.333 \\
		BERTL (EB $\leftarrow$ AIT) \cite{VAD-SOTA} & 0.765 & \textbf{0.583} & 0.416 \\
		\hline
		\hline
		VADR\textsubscript{RoBERTa} & 0.804 & 0.494 & \textbf{0.511} \\
		\quad w/ Empath & 0.798 & 0.482 & 0.510 \\
		\hline
		VADR & 0.821 & 0.553 & 0.493 \\
		\hline  
		VADEC (AIT) & 0.820 & 0.563 & 0.459 \\
		\hline
		VADEC (SenWave) & \textbf{0.823} & 0.553 & 0.485 \\
		\hline
	\end{tabular}
	\end{adjustbox}
	\vspace{-1em}
\end{table}

\subsection{COVID-19 and Indians: A Case Study}
\label{subsec:case_study}
\begin{table*}
	\caption{Few Examples of Single and Multi-label Predictions on Tweets from \textit{Twitter\_IN}}
	\label{tab:prediction_examples}
	\vspace{-1em}
	\centering
	\resizebox{\textwidth}{!}{
	\begin{tabular}{|p{0.75\textwidth}|c|}
		\hline
		\makebox[0.75\textwidth]{\textbf{Tweet}} & \textbf{Predicted Labels} \\
		\hline
		\multicolumn{2}{|l|}{\textbf{Single Label}} \\ 
		\hline
		Let us spare a moment and thought for the junior resident doctors of Mumbai on the frontline fighting it out alone with little help from the government against all odds and at great personal risk & Thankful \\
		\hline
        This is the time to fight Covid19 at present but some intelligent Generals are focusing on war and terrorism & Annoyed \\
        
		\hline
		\multicolumn{2}{|l|}{\textbf{Multiple Labels}} \\ \hline
		The first Covid 19 positive from Meghalaya Dr John Sailo Rintathiang passed away early this morning. Sailo 69 who was also the owner of Bethany hospital was tested positive on April 13 2020 & Sad, Official Report \\
		\hline
		Media is so obsessed with a particular community that they even misspell coronavirus & Annoyed, Joking, Surprise \\

		\hline			
	\end{tabular}}
\end{table*}

\begin{table*}
	\caption{Major aspects affecting various emotions among Indians towards the COVID-19 pandemic.}
	\label{tab:top_aspects}
	\vspace{-1em}
	\centering
	\begin{tabular}{lp{0.8\textwidth}}
		\hline
		\textbf{Emotion} & \textbf{Major aspects} \\ 
		\hline
		
% 		\textbf{Anxious} & family, symptom, test, treatment, rate, risk, mask, spread, zone, assault \\
% 		\textbf{Anxious} & family, symptom, test, treatment, rate, risk, mask, spread, zone, assault, brexit, NHS \\
		
		\textbf{Annoyed} & govt, politics, death, news, religion, jamaat, work, China, assault, border \\
		
		\textbf{Sad} & lockdown, death, distancing, life, family, economy, village, doctor, worker, school \\
		
% 		\textbf{Pessimistic} & price, business, infection, demise, peak, curve, communalism, war, situation, depression \\
        
        \textbf{Thankful} & doctor, service, staff, nurse, app, fund, assistance, leadership \\
		
		\textbf{Optimistic} & initiative, opportunity, measure, arogyasetuapp, IndiaFightsCorona, stayhome, vaccine, change, support, action \\
        \hline		
	\end{tabular}
\end{table*}

\begin{figure*}
    \centering
    \begin{tabular}{cc}
	\includegraphics[width=0.48\textwidth, height=0.27\textwidth]{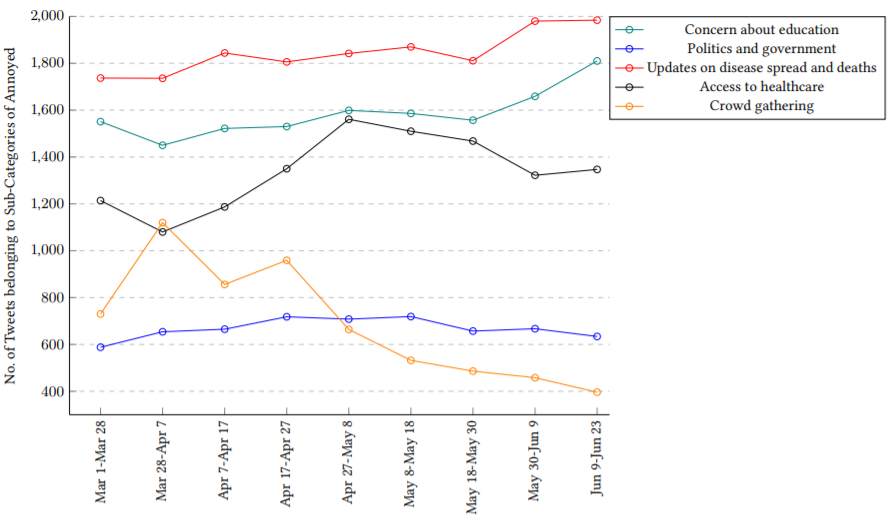} &
	\includegraphics[width=0.48\textwidth, height=0.27\textwidth]{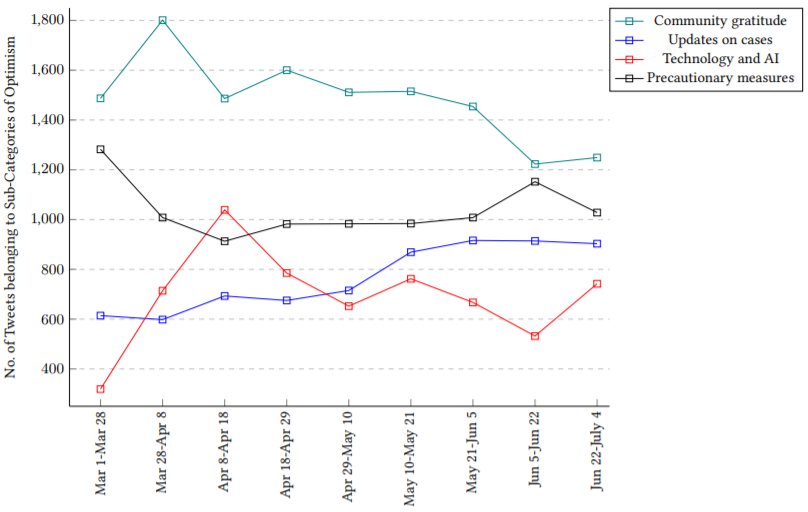} \\
	\textbf{(a) Annoyed} & \textbf{(b) Optimism} \\
	\end{tabular}
	\vspace{-1em}
	\caption{Change in Sub-categories of Emotional Triggers towards the COVID-19 pandemic over time.}
	\label{fig:triggers}
% 	\vspace{-0.5em}
\end{figure*}

For this analysis, we consider \textbf{Twitter\_IN}, a subset of \textit{COVID-19 Twitter chatter} dataset (version 17) \cite{Banda2020ALC}, containing around 140K English tweets from India posted between January 25th and July 4th 2020. Owing to very few reported cases in India before March 2020, we begin our analysis by predicting emotions from tweets, posted on or after Match 1st 2020, using VADEC trained on \textit{EMOBANK} and \textit{SenWave}. Few tweets with their predicted emotions are listed in Table \ref{tab:prediction_examples}. For each emotion, we obtain its contributing aspects by training an unsupervised neural topic model, ABAE (He et al. \cite{He2017AnUN}) on the subset of tweets containing the given emotion as per \textit{VADEC} predictions. Few emotions along with their most accurate aspects are reported in Table \ref{tab:top_aspects}. For each emotion, the extracted aspect terms are further filtered and assigned meaningful sub-categories by means of a many-to-many mapping. In Figure \ref{fig:triggers}, we plot the temporal trends of these sub-categories (with roughly equal-sized bins in terms of no. of tweets predicted with the emotion plotted) that respectively made Indians feel \textit{annoyed} (Fig. \ref{fig:triggers}a) and \textit{optimistic} (Fig. \ref{fig:triggers}b) over time. In Fig. \ref{fig:triggers}a, the peak in \textit{Crowd gathering} between March 28th and April 7th can be attributed to the \textit{Tablighi Jamaat} gatherings\footnote{\url{https://en.wikipedia.org/wiki/2020_Tablighi_Jamaat_COVID-19_hotspot_in_Delhi}} unfortunately triggering widespread criticism. Fig. \ref{fig:triggers}b shows a high level of \textit{Community gratitude} in general, with occasional peaks which may be attributed to the events targeted at raising solidarity among the public. For \textit{Technology and AI}, we observe a peak near the launch date of the \textit{Arogya Setu App}\footnote{\url{https://en.wikipedia.org/wiki/Aarogya_Setu}} - developed by the Indian Government to identify COVID-19 clusters.
\section{Conclusion and Future Work}\label{conclusion}
In this work, we for the first time exploit the correlation between \textit{categorical} and \textit{dimensional} models of emotion analysis by proposing \textit{VADEC}, a multi-task affect classifier with the primary objective of efficiently detecting emotions from tweets. Co-training the tasks of \textit{multi-label emotion classification} and \textit{multi-dimensional emotion regression} helps the former thereby achieving state-of-the-art results on two benchmark datasets, \textit{AIT} (non-COVID) and \textit{SenWave} (COVID-related). For the \textit{regression} task, \textit{VADEC} still outperforms the strongest baseline on the \textit{EMOBANK} dataset on the \textit{V} and \textit{D} dimensions. In future, we would like to investigate the hierarchical relationship between the tasks and analyze the relative impact of each emotion dimension on the emotion classification task.\if 0 We conclude by detecting and analyzing the contributing factors for various emotions among Indians towards the COVID-19 pandemic from their tweets.\fi

\begin{acks}
    This research is partially supported by IMPRINT-2, a national initiative of the Ministry of Human Resource Development (MHRD), India. Niloy Ganguly was partially funded by the Federal Ministry of Education and Research (BMBF), Germany (grant no. 01DD20003).
\end{acks}

\newpage
\bibliographystyle{ACM-Reference-Format}
\bibliography{main}

\end{document}